\documentclass[journal=jacsat,manuscript=article]{achemso}

\usepackage{chemformula} 
\usepackage[T1]{fontenc} 
\usepackage[
]{achemso}
\setkeys{acs}{maxauthors=10}
\setkeys{acs}{etalmode=truncate}

\usepackage{tabu}
\usepackage[printonlyused]{acronym}
\usepackage{multirow}

\usepackage{changes}

\usepackage{graphicx}
\usepackage{dcolumn}
\usepackage{bm}
\usepackage{physics}
\usepackage{braket}
\usepackage{afterpage}
\newcommand{\tj}[6]{ \begin{pmatrix}
  #1 & #2 & #3 \\
  #4 & #5 & #6
\end{pmatrix}}

\author{Dominik Ertel*}
\affiliation{Physikalisches Institut, Albert-Ludwigs-Universit\"{a}t Freiburg, Hermann-Herder-Stra{\ss}e 3, 79104 Freiburg, Germany.}
\author{David Busto*}
\affiliation{Physikalisches Institut, Albert-Ludwigs-Universit\"{a}t Freiburg, Hermann-Herder-Stra{\ss}e 3, 79104 Freiburg, Germany.}
\alsoaffiliation{Department of Physics, Lund University, PO Box 118, SE-221 00 Lund, Sweden.}
\author{Ioannis Makos}
\affiliation{Physikalisches Institut, Albert-Ludwigs-Universit\"{a}t Freiburg, Hermann-Herder-Stra{\ss}e 3, 79104 Freiburg, Germany.}
\author{Marvin Schmoll}
\affiliation{Physikalisches Institut, Albert-Ludwigs-Universit\"{a}t Freiburg, Hermann-Herder-Stra{\ss}e 3, 79104 Freiburg, Germany.}
\author{Jakub Benda}
\affiliation{Institute of Theoretical Physics, Faculty of Mathematics and Physics, Charles University, V Holešovičkách 2, 180 00, Prague 8, Czech Republic.}
\author{Francesca Bragheri}
\affiliation{Istituto di Fotonica e Nanotecnologie, CNR, 20129 Milano, Italy.}
\author{Roberto Osellame}
\affiliation{Istituto di Fotonica e Nanotecnologie, CNR, 20129 Milano, Italy.}
\author{Eva Lindroth}
\affiliation{Department of Physics, Stockholm University, AlbaNova University Center, SE-106 91 Stockholm, Sweden.}
\author{Serguei Patchkovskii}
\affiliation{Max Born Institute, Max-Born-Str. 2A, D-12489 Berlin, Germany.}
\author{Zdeněk Mašín}
\affiliation{Institute of Theoretical Physics, Faculty of Mathematics and Physics, Charles University, V Holešovičkách 2, 180 00, Prague 8, Czech Republic.}
\email{zdenek.masin@matfyz.cuni.cz}
\author{Giuseppe Sansone}
\affiliation{Physikalisches Institut, Albert-Ludwigs-Universit\"{a}t Freiburg, Hermann-Herder-Stra{\ss}e 3, 79104 Freiburg, Germany.}
\email{giuseppe.sansone@physik.uni-freiburg.de}

\title[An \textsf{achemso} demo]
  {Anisotropy parameters for two-color photoionization phases in randomly oriented molecules: theory and experiment in methane and deuteromethane.}

\abbreviations{IR,NMR,UV}

\begin{document}
\footnotetext{* They contributed equally to this work}
%
%
%
%
%

\begin{abstract}
We present a combined theoretical and experimental work investigating the angle-resolved phases of the photoionization process driven by a two-color field consisting of an attosecond pulse train and an infrared pulse in an ensemble of randomly oriented molecules. We derive a general form for the two-color photoelectron (and time-delay) angular distribution valid also in the case of chiral molecules and when relative polarizations of the photons contributing to the attosecond photoelectron interferometer differ. We show a comparison between the experimental data and the theoretical predictions in an ensemble of methane and deuteromethane molecules, discussing the effect of nuclear dynamics on the photoionization phases. Finally, we demonstrate that the oscillating component and the phase of the two-color signal can be fitted using complex asymmetry parameters, in perfect analogy with the atomic case.
\end{abstract}

\section{Introduction}
The availability of isolated attosecond pulses and their trains~\cite{RMP-Krausz-2009} has enabled the investigation of purely electronic processes occurring on a comparable timescale~\cite{Calegari2016, CPC-Sansone-2012}. Among several processes, the photoionization mechanism has been widely investigated in the last years using attosecond technology to derive information about the influence of the electronic structure on the emission of photoelectron wave packets after the absorption of an extreme ultraviolet (XUV) photon. In particular, the influence of electronic correlation in atoms~\cite{Isinger2017}, the role played by shape resonances in molecules~\cite{Nandi2020} and the effect of an anisotropic molecular potential on the photoemission phases~\cite{Ahmadi2022,Vos2018a} have been elucidated through attosecond photoelectron interferometry based on the combination of XUV attosecond pulse trains and a synchronized infrared (IR) field~\cite{Science-Paul-2001}. When combined with photoelectron spectrometers with angular resolution, this technique, usually indicated as reconstruction of attosecond beating by interference of two-photon transitions (RABBIT), allows one to retrieve three-dimensional momentum information about the emitted photoelectrons, giving access, for example, to the angular dependence of the photoemission phases or time delays. In atoms, this dependence has been studied for the case of non-resonant photoionization, demonstrating the presence of a minimum in the direction perpendicular to the common polarization direction of the XUV and IR fields~\cite{Heuser2016}. Moreover, the angular-resolved information gives the possibility of realizing a complete experiment, with the determination of the amplitudes and relative phases of the partial waves involved in the photoionization process~\cite{NATCOMM-Peschel-2022}. The flexibility offered by the RABBIT approach, in which the polarization state and directions of the XUV and IR fields can be independently controlled, has been exploited for the investigation of attosecond circular dichroism in two-color photoionization in atoms~\cite{NATPHYS-Han-2023}.

In molecular systems, the availability of information about the emission direction of the photoelectron wave packet was used to elucidate the effect of the molecular potential on the outgoing electron and to reconstruct attosecond time delays in the recoil frame~\cite{Biswas2020, SCIADV-Heck-2021,Ahmadi2022}.

As demonstrated in ref.~\cite{PRA-Lindroth-2021}, in atoms, the angular dependence of the photoionization phases and time delays can be parametrized 
using two complex asymmetry parameters $\tilde\beta_2$ and $\tilde\beta_4$, in analogy to the well-known formula describing the photoelectron angular distributions $I(\theta)$ in a two-color field~\cite{Reid2003}:
\begin{equation}\label{Eq1}
  I(\theta)=I_0\left[1+\beta_2P_2(\cos\theta)+\beta_4P_4(\cos\theta)\right]
\end{equation}
where $\theta$ indicates the photoemission angle with respect to the common polarization of the XUV and IR fields, $\beta_2$ and $\beta_4$ are (real) asymmetry parameters for a two-photon ionization process, and $P_2$ and $P_4$ are the Legendre polynomials of the second and fourth order.

In this work, we demonstrate that the complex parametrization of the photoionization phases can be extended to the case of an ensemble of randomly oriented molecules in the laboratory frame (LF). This finding is consistent with the observation that the photoelectron angular distributions (PADs) generated in a molecular ensemble in similar experimental conditions are described by an expression analogous to Eq.~(\ref{Eq1})~\cite{Reid2003}. We present a general derivation of the interfering terms involved in the RABBIT process, considering also the case of chiral systems. Finally, we apply the formalism to the angle-resolved photoionization phases measured in a randomly oriented ensemble of \ch{CH4} and \ch{CD4} molecules, investigating the PADs and photoionization phases associated with final non-dissociating states of the molecular cations (\ch{CH4+} and \ch{CD4+}).

\section{Methods}
\subsection{Experimental setup}
\begin{figure}[h!]
\centering 
\includegraphics[scale=1]{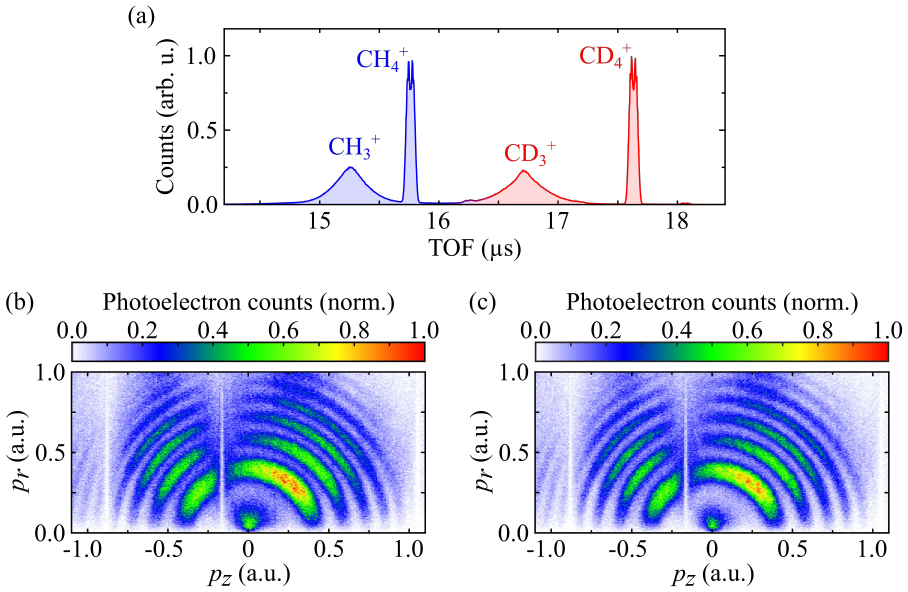}
  \caption{(a) XUV-only experimental TOF mass spectrum acquired in an equal mixture of \ch{CH4} and \ch{CD4} molecules. XUV-only photoelectron angular distributions measured in coincidence with the ionic fragments \ch{CH4+} (b) and \ch{CD4+} (c). The laser polarization is along the horizontal axis. Adapted with permission from ref.~\cite{PHDErtel2022}~.  Copyright 2022 University of Freiburg.} 
  \label{Fig1}
\end{figure}

Measurements were conducted in an equal mixture of \ch{CH4} and \ch{CD4} molecules utilizing a photoelectron-photoion coincidence spectrometer (Reaction Microscope)~\cite{Dorner2000, Ullrich2003}.
XUV photons in the spectral range between 10~eV- and 50~eV were delivered by a source of attosecond pulse trains operating at 50~kHz~\cite{RSI-Ertel-2023}. The sidebands of the main photoelectron lines were generated by a synchronized IR pulse with an estimated intensity of I$_{\mathrm{IR}}$$\approx 5\times 10^{11}$~W/cm$^2$. The delay between the XUV and IR pulse was varied using a system of drilled glass plates in a collinear configuration \cite{JP-Ahmadi-2020, RSI-Ertel-2023}.
Figure~\ref{Fig1}a presents the typical time-of-flight (TOF) mass spectrum generated by single XUV-photoionization in the \ch{CH4}-\ch{CD4} mixture. The TOF is characterized by dissociating (\ch{CH3+} and \ch{CD3+}) and non-dissociating channels (\ch{CH4+} and \ch{CD4+}), in agreement with spectroscopic information obtained with monochromatic XUV radiation \cite{Chupka1968, Field1995, Latimer1999}. The presence of a second dissociating channel (\ch{CH2+} and \ch{CD2+}) could not be identified due to the low branching ratio of this channel in the photon energy range covered by our XUV source.
The PADs associated with the non-dissociating channels are shown in Fig.~\ref{Fig1}b and c for \ch{CH4+} and \ch{CD4+}, respectively. The plots are presented in the plane $p_r-p_z$ corresponding to the components of the photoelectron momenta $\mathbf{p}$ perpendicular ($p_r$) and parallel ($p_z$) to the spectrometer axis. The polarization of the XUV radiation was oriented along the spectrometer axis.
The PADs are characterized by photoelectron peaks corresponding to the absorption of a single XUV-photon. The width of a single photoelectron peak is broader than the bandwidth of the corresponding XUV harmonic ($\approx$ 200~meV (full-width at half maximum; FWHM)), due to the finite resolution of the photoelectron spectrometer
and the energy width of the vibronic absorption band of methane and
deuteromethane \cite{Field1995}.
In Fig. 1b and 1c, the sharp vertical artefacts indicate an information loss about the photoelectrons momenta, known as magnetic nodes. This loss occurs when the photoelectrons complete an exact integer multiple of cycles in their cyclotron motion towards the detector. Therefore, to ensure the integrity of our results, the analysis considers only experimental data that exhibits positive momentum $p_z$ values, i.e., photoemission angles up to $90^{\circ}$ with respect to the spectrometer axis.

For the two-color photoionization experiments, the IR and XUV pulses shared the same polarization direction. The relative delay between the two fields was changed in steps of $\Delta t\approx 200$~as. Sideband oscillations were acquired in a range of 12 fs and the total acquisition time was about 96 hours.

\subsection{Theory}
In this section, we derive expressions for the complex $\tilde\beta$ parameters which fully characterize the angular dependency of the time delays. The topic of PAD in both single- and multi-photon ionization has been extensively studied both theoretically and experimentally, see e.g. the review by Reid~\cite{Reid2003} and the references therein.
The general expression for $n$-photon PAD for randomly oriented molecules was presented earlier~\cite{benda2021}.
While the result presented there was for the direct $n$-photon contribution, the formula remains identical for the interference term too if the product of the partial-wave matrix elements is replaced with the product of dipoles coming from the two interfering pathways. By itself, this expression is sufficient to show that the atomic and molecular two-photon delays can be parametrized equivalently for the simplest case of XUV and IR fields linearly polarized in the same direction and for non-chiral molecules. Here, we outline the theory for the special case of interference of two-photon transitions. In Appendices A and B we perform the derivation in detail and analyze the resulting expression further for the case of various polarizations of the two fields including the case of chiral molecules (or atoms prepared in chiral states) (see Appendix A). To reduce the complexity we specialize the discussion to co-propagating XUV and IR pulses. In the following, we use Hartree atomic units ($m=e=\hbar=1$) unless noted otherwise.

The two-photon photoionization matrix elements employed here in applications of the theory were calculated using the molecular R-matrix codes UKRMol+\cite{CPC-Masin-2020}. The details of the calculations were described previously~\cite{SCIADV-Ertel-2023}.

\subsection{Photoelectron angular distributions in the RABBIT process}

\begin{figure}[!ht]
    \centering
    \includegraphics[width=0.8\textwidth]{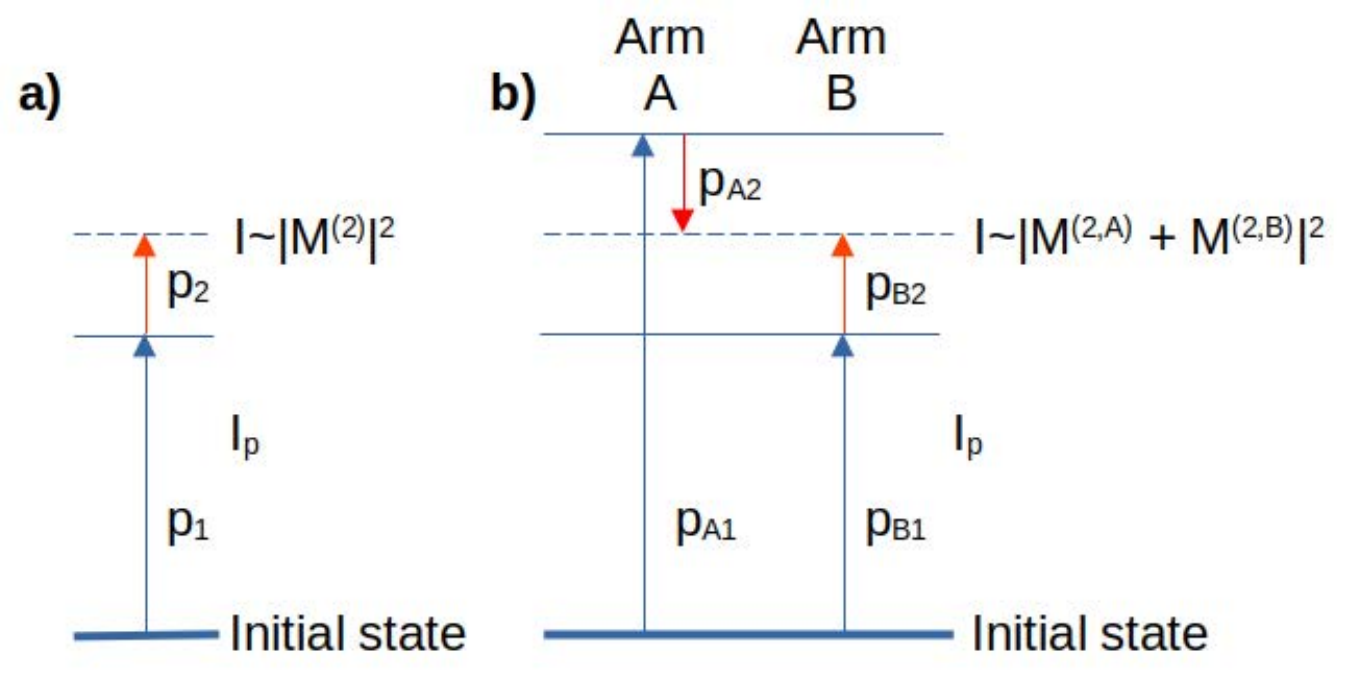}
    \caption{a) Direct two-photon above-threshold ionization. b) The RABBIT interferometer comprises two arms A and B corresponding to the two-photon amplitudes $M^{(2,A)}$ and $M^{(2,B)}$ which combine to produce the sideband signal $I^{(p_{A1},p_{B1}),(p_{A2},p_{B2})}$. Each arm contains two photons with polarizations $p_{X1}$, $p_{X2}$, where $X=A,B$. $I_{p}$ stands for the ionization potential of the molecule.}
    \label{Fig2}
\end{figure}

The momentum space two-photon matrix element has the form
\begin{eqnarray}
    M_{fi}^{(2)} = \langle \Psi_{f\bm{k}}^{(-)} | \hat{D}_2 \hat{G}_1^{(+)} \hat{D}_1 | \Psi_i \rangle,\label{eq:2pdef}
\end{eqnarray}
where $\ket{\Psi_{f\bm{k}}^{(-)}}$ is the final continuum wavefunction, $\hat{D}_1$ and $\hat{D}_2$ are the dipole operators, $\ket{\Psi_i}$ is the initial bound state wavefunction and $\hat{G}_1^{(+)}$ is the field-free Green's function of the system which generates the intermediate state at energy $E_{i}+\omega$ (or $E_{i}-\omega$, depending on the arm of the interferometer, see Fig.~\ref{Fig2}) after absorption of the XUV photon ($\omega$ indicates the central frequency of the IR field). A detailed description of the calculation of this matrix element using the R-matrix method is given in our previous work~\cite{benda2021}. The RABBIT signal is given by the interference term originating from two two-photon pathways, see Fig.~\ref{Fig2}b, leading to the same final continuum state:
\begin{eqnarray}\label{eq:signal}
    I \sim | M_{fi}^{(2,A)} + M_{fi}^{(2,B)}|^2 = | M_{fi}^{(2,A)} |^2 + |M_{fi}^{(2,B)}|^2 + \mathrm{Re}[ 2|M_{fi}^{(2,A)}| |M_{fi}^{(2,B)}|e^{\mathrm{i} (\phi_{>} - \phi_{<})}e^{\mathrm{i} (\Phi_{>}-\Phi_{<})}],
\end{eqnarray}
where $\Phi_{<}$ and $\Phi_{>}$, respectively the phases of the XUV spectral components in the absorption and emission path, determine the relative timing of the XUV and IR fields. The phases $\phi_{<}$ and $\phi_{>}$ are the intrinsic phases of the involved two-photon matrix elements.
The molecular-frame RABBIT delay is then given by the phase of the exponential, i.e.
\begin{eqnarray}\label{eq:RABBITT_MF}
    \tau^{2\omega}(\theta_{MF},\phi_{MF}) = \frac{(\phi_{>} - \phi_{<})}{2\omega} = \frac{\mathrm{arg}[M_{fi}^{(2,A)}M_{fi}^{(2,B)*}]}{2\omega}.
\end{eqnarray}
Here the angles $\theta_{MF},\phi_{MF}$ specify the direction of the photoelectron momentum in the molecular frame.
Orientationally-averaged delays are computed from the phase of the orientationally-averaged interference term written in the LF, where the photoelectron momentum $\bm{k}$ and photon polarizations are defined:
\begin{eqnarray}
    \tau^{2\omega}_{Av}(\theta,\phi) &=& \frac{\mathrm{arg}[I^{(p_{A1},p_{B1}),(p_{A2},p_{B2})}]}{2\omega},\label{eq:intf}\\
    I^{(p_{A1},p_{B1}),(p_{A2},p_{B2})} &=& \frac{1}{8\pi^2}\int\mathrm{d}R M_{fi}^{(2,A)}(R)\left(M_{fi}^{(2,B)}(R)\right)^{*},
\end{eqnarray}
where the angles $\theta,\phi$ indicate the direction of the photoelectron momentum in the LF. The integration is over the three Euler angles which rotate the molecule from the molecular frame to the laboratory frame and $p_{A1}$, $p_{A2}$ and $p_{B1}$, $p_{B2}$ denote polarizations of the first and the second photon in the spherical basis in the LF as shown in Fig.~\ref{Fig2}. 
 We note that in the case of orientational (or vibrational, see below) averaging, the individual phases $\phi_{>}$ and $\phi_{<}$ are no longer well-defined, unlike in the molecular-frame case. Their difference, given by the phase of $I^{(p_{A1},p_{B1}),(p_{A2},p_{B2})}$, is however sufficient to define the observables.

The final result expressed in the spherical basis for the four-photon polarizations is
\begin{eqnarray}
&&I^{(p_{A1},p_{B1}),(p_{A2},p_{B2})}=\sum_{L=0}^{4}b_{L,M}Y_{L,M}^{*}(\bm{k}),\label{eq:averaged_intf}\\
&&M = p_{B1}+p_{B2}-(p_{A1}+p_{A2}),
\end{eqnarray}
where $b_L$ are coefficients whose explicit form is given in Eq.~\eqref{eq:averaged_main}. In most RABBIT experiments to date all four photon polarizations are linear and equal implying $M=0$. Properties of the various other schemes employing different photon polarizations are discussed in Appendix A and the complete derivation leading up to Eq.~\eqref{eq:averaged_intf} can be found in Appendix B.

\subsubsection{Direct two-photon ionization vs RABBIT}

The scheme for the direct two-photon ionization process is compared in Fig.~\ref{Fig2} to the RABBIT case. As shown in Appendix A the direct two-photon photoelectron angular distribution will have the same shape as in RABBIT when equal polarizations in the two arms of the interferometer are used, cf. Eq~\eqref{Eq1}. Terms with non-zero $M$ appear in the skewed-polarized case in both RABBIT~\cite{jiang2022a} and direct two-photon ionization~\cite{Cacelli1991, reid1991}. However, as shown in Appendix A (see Eq.~\eqref{eq:M_val}), only in the case of RABBIT we have (at least in theory) the possibility to control precisely the exact $M$ value of the contributing spherical harmonic by fixing the polarizations of the four photons independently. Furthermore, in RABBIT the quantum number $M$ can assume values such that $|M| \leq 4$, whereas in the direct case it is  limited to $|M| \leq 2$. In this sense, RABBIT is a more selective probe of photoionization dynamics compared to direct two-photon ionization. Nevertheless, even the usual RABBIT setup with equal polarizations is more informative than the direct process since it provides access not only to the PAD but also to the photoionization phase. 

In this case, due to the interference between the two photoionization pathways involving consecutive harmonics, the signal of the sideband (SB) of order $N$ is described by the relation:
\begin{equation}
    I^{SB}_{N}(\Delta t;E)=A_{0\omega}+A_{2\omega}\cos(2\omega\Delta t-\Delta\varphi),
    \label{Eq2}
\end{equation}
where 
the parameters $A_{0\omega}$, $A_{2\omega}$ and $\Delta\varphi$ indicate the constant and oscillating component of the RABBIT signal and the phase of its oscillation. In our analysis, these three parameters depend on the photoelectron energy $E$ and on the angle $\theta$, while $\Delta t$ indicates the relative delay between the attosecond pulse train and the IR field. The modulus of the interfering term corresponds to the amplitude of the oscillating component of the sideband signal $A_{2\omega}$ and is proportional to $I^{p_{1},p_{2}}$, a particular case of the general form introduced in Eq.~\eqref{eq:averaged_intf} whose expression is reported in Eq.~\eqref{eq:averaged}. Moreover, the phase $\Delta\varphi$ is equal to the numerator reported in Eq.~\eqref{eq:intf}.

For randomly oriented molecules, in analogy to atoms, the angular dependence of the interference contribution to the sideband signal can be parametrized using complex beta parameters $\tilde\beta_2$ and $\tilde\beta_4$ (see Eq.~\eqref{eq:complexbeta} in Appendix B).
As shown later in the case of the experiments in methane and deuteromethane, these quantities can be extracted 
using the relations:
\begin{eqnarray}\label{Eq_fitting1}
    A_{2\omega}(\theta)=\left|\frac{A_{int}}{4\pi}\left(1+\sum_{L=2,4}\tilde\beta_LP_L(\cos\theta)\right)\right|\,,\\\label{Eq_fitting2}
    \Delta\varphi(\theta)=\mathrm{arg}\left[\frac{A_{int}}{4\pi}\left(1+\sum_{L=2,4}\tilde\beta_LP_L(\cos\theta)\right)\right]\,,
\end{eqnarray}
where $A_{int}$ describes the total oscillating component $A_{2\omega}$ integrated over all angles.

\subsection{Treatment of nuclear-motion effects and spectral simulations}

The theory of nuclear-motion effects in RABBIT spectra and the numerical details of the simulations in methane have been described elsewhere\cite{PRA-Patchkovskii-2022,SCIADV-Ertel-2023}, and need not be fully restated here. Briefly, the nuclear-motion effects encompass three distinct, isotope-specific contributions: the change in the average geometry upon ionization, vibronic energy redistribution in the cation, and the coherent averaging over the spatial extent of the zero-point vibrational wavefunction.

The average-geometry contribution is treated by considering electronic matrix elements evaluated at the ``characteristic'' nuclear geometry, arising due to the short-time dynamics on the cationic potential-energy surfaces upon photoionization. For \ch{CH4} (\ch{CD4}) these geometries are taken $1.17$~fs ($1.27$~fs) after the ionization event\cite{SCIADV-Ertel-2023}. They exhibit $D_{2d}$ symmetry, lower than the nominal $T_d$ symmetry of the equilibrium neutral species. The effects of the vibronic-energy redistribution, leading to the finite energy width of the photoelectron spectra, are treated within the autocorrelation formalism\cite{PRA-Patchkovskii-2022}. These effects enter the electronic matrix elements via an effective ionization potential, at each photoelectron energy and RABBIT sideband.

The zero-point effects are treated in the harmonic approximation for the nuclear vibrational wavefunctions and second-order finite-difference expansion of the electronic matrix elements around the high-symmetry initial geometry. Due to the high cost of the matrix-element evaluation, the (small) zero-point correction is only evaluated at one photoelectron energy within each sideband, corresponding to the vertical IP in the valence photoionization band ($14.4$~eV).

For numerical simplicity, all nuclear-motion effects are evaluated in the molecular frame, and then averaged over the molecular orientations numerically, using the order-17 Lebedev grid for the $\alpha,\beta$ Euler angles and a uniform grid of the matching order for the $\gamma$ Euler angle. The complex $\tilde\beta$ coefficients were then determined by ``observing'' the photoelectrons at three angles relative to the common polarization direction of the laser fields, corresponding to the zeros of the $P_2$ and $P_4$ Legendre polynomials and fitting the Eq.~\eqref{Eq_fitting1} to the results. We have verified that the fit reproduces the calculated profile at other photoelectron-detection angles exactly. In the absence of nuclear-motion effects, the numerical procedure agrees exactly with the analytical result given by the Eq.~\eqref{eq:averaged_intf}. The numerical results are illustrated below, in Fig.~\ref{Fig3}.

\begin{figure}
\centering \resizebox{1.0\hsize}{!}{
\includegraphics{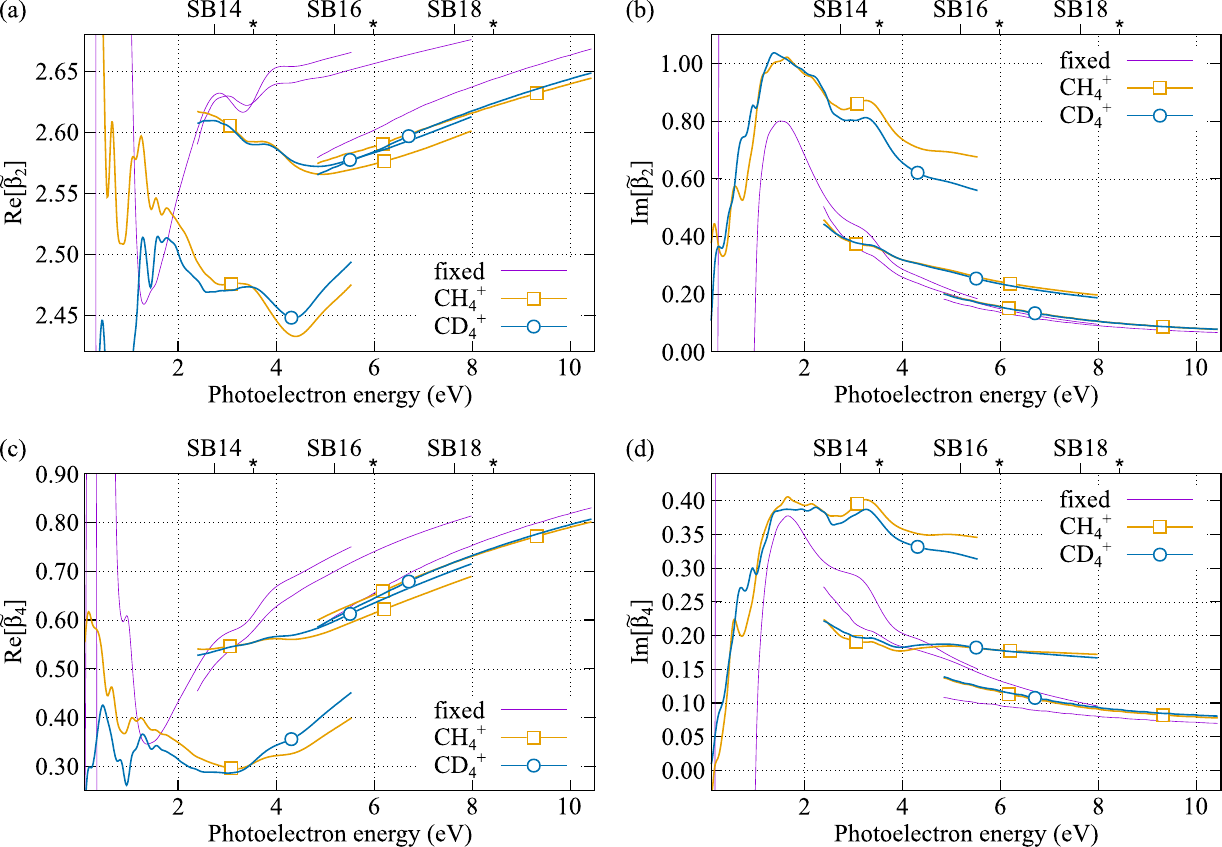}}
\caption{Calculated asymmetry parameters as a function of photoelectron energy. On all panels, the lower horizontal axis gives photoelectron energy in eV. The upper horizontal axis indicates the side-band number; the position of the maximum of the photoelectron peaks correlating to non-fragmenting \ch{CH4+}/\ch{CD4+} photoions (IP$\approx13.61$ eV) within each sideband are indicated with asterisks. Magenta lines indicate values calculated for the fixed nuclei at the equilibrium \ch{CH4} geometry. Gold lines with open squares (teal lines with open circles) show the results including nuclear motion effects in \ch{CH4} (\ch{CD4}). The panels are: the real (a) and imaginary (b) part of $\tilde{\beta_2}$ and the real (c) and imaginary (d) part of $\tilde{\beta_4}$ coefficient. See text for the numerical details.}
  \label{Fig3}
\end{figure}

For the comparison with the experimental results presented in the following sections, the theoretical predicted oscillating component, phase, and complex asymmetry parameters are taken at the center of the associated ionic (\ch{CH4+} or \ch{CD4+}) band.

\section{Results and discussion}

\subsection{XUV photoionization of CH$_4$ and CD$_4$}
\begin{figure}
\includegraphics{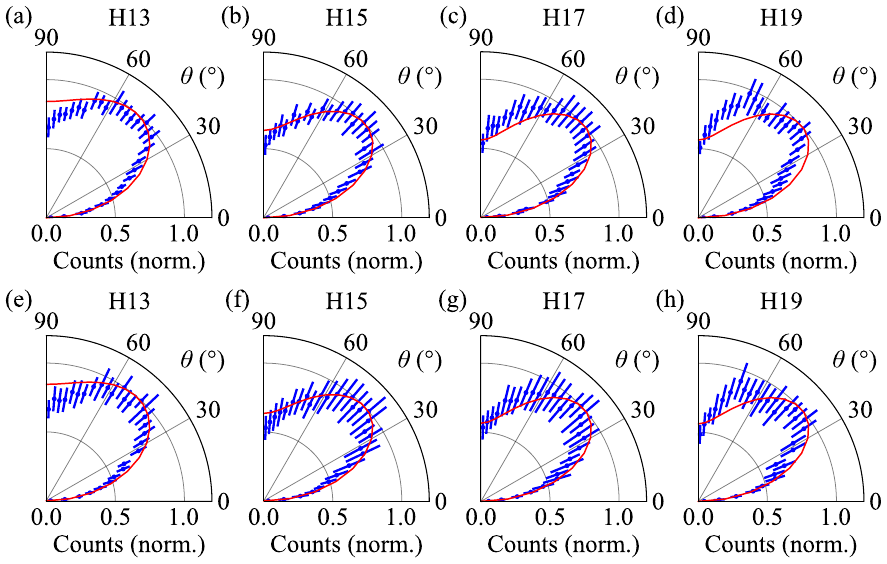}
  \caption{Experimental LF PADs (blue points and error bars) measured in coincidence with the ionic channels \ch{CH4+} (a-d, upper row) and  \ch{CD4+} (e-h, lower row) using a train of attosecond pulses consisting of the odd harmonics of the fundamental radiation H13 (a,e), H15 (b,f), H17 (c,g), H19 (d,h). The error bars are the standard deviations within the energy integration range. The theoretical predictions, multiplied by $\sin(\theta)$ in order to take into account the geometrical effect related to the solid angle of the detector, are presented in red. Adapted with permission from ref.~\cite{PHDErtel2022}~. Copyright 2022 University of Freiburg.}
  \label{Fig4}
\end{figure}

\begin{figure}
\includegraphics{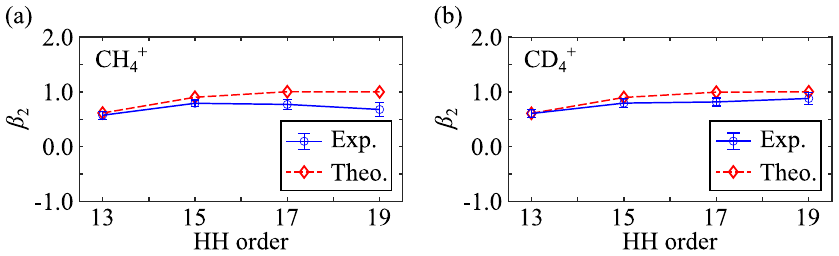}
  \caption{Comparison of the $\beta_2$ asymmetry parameter for different harmonic orders derived from the fitting of the LF PADs (blue open circles and solid lines) and from the R-matrix calculations~\cite{SCIADV-Ertel-2023} using UKRmol+ (red diamonds and dashed lines) for the PADs measured in coincidence with the ionic channels \ch{CH4+} (a) and \ch{CD4+} (b). The error bars were determined from the fitting procedure of the PADs. Adapted with permission from ref.~\cite{PHDErtel2022}~. Copyright 2022 University of Freiburg.}
  \label{Fig5}
\end{figure}

Figure~\ref{Fig4} reports the comparison between the LF PADs measured in the XUV-only case and the theoretical predictions for the energies corresponding to the absorption of different harmonics orders from H13 to H19. The PADs measured in coincidence with \ch{CH4+} and \ch{CD4+} are shown in the upper (a-d) and lower (e-h) panels, respectively. The agreement between the experiment and theoretical predictions (shown in red) is excellent. 
For high kinetic energy photoelectrons ((H19, in Fig.~\ref{Fig4}d and h), small discrepancies above 60$^{\circ}$can be observed. This might originate from a combination of low statistics and higher sensitivity of high energy photoelectrons to magnetic nodes.
The quality of the agreement is further supported by the comparison between the $\beta_2$ parameter extracted from the fitting procedure of the experimental and theoretical data reported in Fig.~\ref{Fig5}a and b for the \ch{CH4+} and \ch{CD4+} ions, respectively.

\subsection{Two-color photoionization of CH$_4$ and CD$_4$}
\subsubsection{Angle-resolved oscillating component and phase of the RABBIT signal}

It was not possible to extract the $\beta_2$ and $\beta_4$ parameters for the two-color photoionization signal directly from the PADs of the sideband signal, due to the large overlap between the single and two-color signal. To isolate the latter term, we first extracted the oscillating component of the photoelectron signal in the energy region centered around the sideband signal by performing either a Fourier Transform and considering the term at frequency $2\omega$ or a fit according to the Eq.~(\ref{Eq2}). We have verified that both two methods deliver consistent results. 

In general, even though the single-photon (XUV) and the two-photon (XUV-IR) paths can lead to the same final photoelectron energy, the absorption of the initial XUV photon along the two pathways will lead to a different final ionic state, thus preventing the observation of an interference effect. Moreover, any possible interference term contributed by a single-photon and a two-photon pathway is expected to wash out when averaging over several pulses, due to the shot-to-shot variation of the carrier-envelope phase of the driving source used in the experiment.

\begin{figure}[h!]
\includegraphics{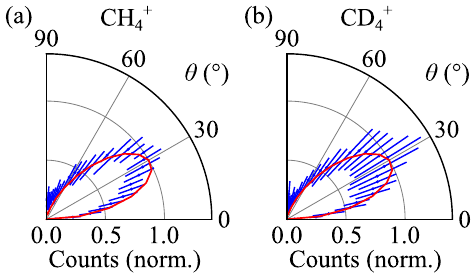}
\caption{Experimental LF angular dependence of the $A_{2\omega}$ components of the photoelectron spectrum corresponding to the SB16 (blue points and error bars) measured in coincidence with the ionic channels \ch{CH4+} (a) and  \ch{CD4+} (b). The error bars were determined as the standard deviation within the energy integration interval. The red curves were obtained from the theoretical predictions following the same procedure adopted for the experimental data. Adapted with permission from ref.~\cite{PHDErtel2022}~. Copyright 2022 University of Freiburg.}
  \label{Fig6}
\end{figure}

\begin{figure}[h!]
\centering 
\includegraphics[scale=1]{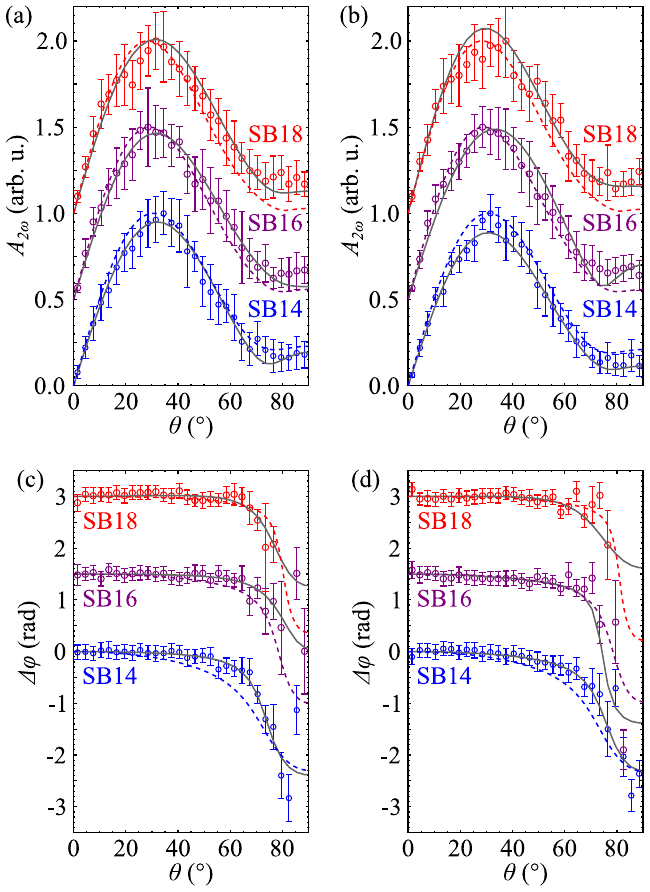}
\caption{Experimental (open circles and error bars) and theoretical (dashed lines) angular dependence of the $A_{2\omega}$ components of the photoelectron spectrum and the sideband phases $\Delta\varphi$ measured in coincidence with the ionic channels \ch{CH4+} (a, c) and \ch{CD4+} (b, d) for the sidebands SB14 (blue), SB16 (purple), and SB18 (red). The theory values are taken at the center of the \ch{CH4+}/\ch{CD4+}-associated band, with the nuclear motion treated as described in ref.\cite{SCIADV-Ertel-2023}. The grey solid lines correspond to the fitting of the distributions obtained using Eqs.~(\ref{Eq_fitting1}, \ref{Eq_fitting2}). Adapted with permission from ref.~\cite{PHDErtel2022}~. Copyright 2022 University of Freiburg.}
  \label{Fig7}
\end{figure}

Figure~\ref{Fig6} shows the angular dependence of the oscillating component $A_{2\omega}(\theta)$ integrated over the energy range of the SB16 (blue symbols and error bars), together with the corresponding quantity extracted from the theoretical simulations for the ions \ch{CH4+} (a) and \ch{CD4+} (b) using the Fourier transform of the photoelectron signal. The agreement in the two-color case is also excellent.
A similar agreement was obtained with the direct fitting of the sideband oscillations using Eq.~(\ref{Eq2}), as presented in Fig.~\ref{Fig7}, which reports the angular dependence of the $A_{2\omega}$ components for \ch{CH4+} (a) and \ch{CD4+} (b). The curves corresponding to different sidebands have been vertically shifted for visual clarity. 
The phase of the sideband oscillations, obtained additionally through the fitting procedure, are depicted in Fig.~\ref{Fig7}c,d for \ch{CH4+} and \ch{CD4+}, respectively. Here, we subtracted from the sideband phase $\Delta\varphi$, introduced in Eq.~(\ref{Eq2}), the phase value obtained around $\theta=0^{\circ}$ (i.e., $\Delta\varphi(\theta=0^{\circ})$) for each sideband.
%
Using this approach, a comparison between the absolute phases of different sidebands is not possible, but it makes the investigation of the angular dependence of the phases and the comparison between consecutive sidebands easier. Moreover, this also removes the influence of the attochirp, which adds a different angle-independent phase offset to the different sidebands.

The comparison between experimental and theoretical data shows, in general, excellent agreement in the interval from $0^{\circ}$ to $70^{\circ}$ (see Fig.~\ref{Fig7}).
The low experimental sideband signal for larger angles makes a reliable determination of the phases of the sideband oscillations around $90^{\circ}$ difficult. Nevertheless, the data indicate the presence of a minimum at this angle for all sidebands. The presence of a minimum for the two-color photoionization phase in the direction perpendicular to the laser polarization in the randomly oriented ensemble of methane and deuteromethane is consistent with the angular dependence of the photoemission phase in two-color fields observed in atoms~\cite{Heuser2016}.

\subsubsection{Parametrization of angle-resolved sideband phases}

We retrieved the complex beta parameters $\tilde\beta_2$ and $\tilde\beta_4$, by performing a simultaneous fit of the oscillating component $A_{2\omega}$ and phase $\Delta\varphi$ according to Eqs.~(\ref{Eq_fitting1},\ref{Eq_fitting2}). The fits of the experimental data are presented in Fig.~\ref{Fig7}, indicating that the parametrization can reproduce qualitatively very well the angular evolution of both the oscillating components and phases.

\begin{figure}
\centering \resizebox{1.0\hsize}{!}{
\includegraphics{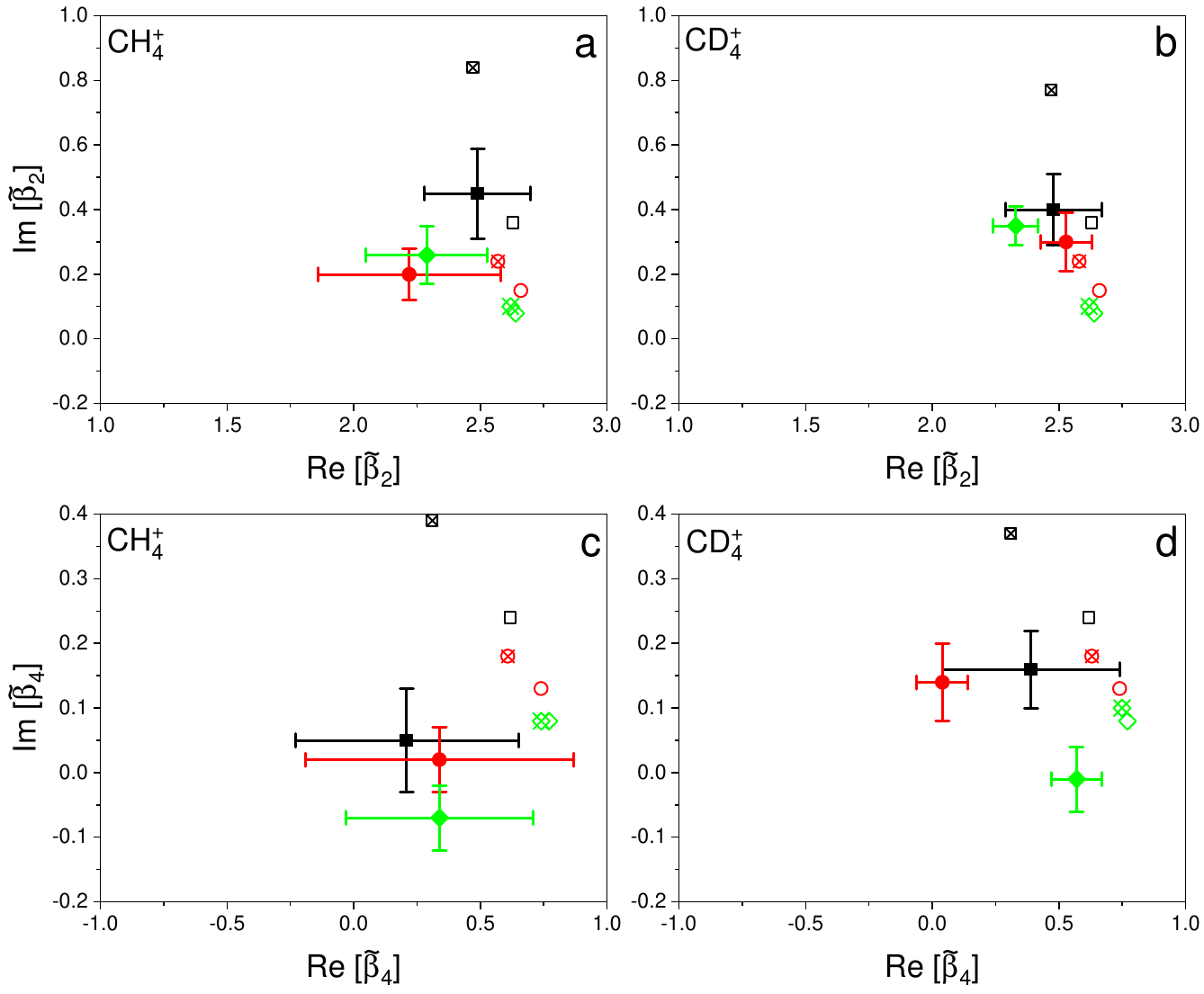}}
\caption{Comparison of the real and imaginary parts of the $\tilde\beta_2$ (a,b) and  $\tilde\beta_4$ (c,d) parameters derived from the experimental data (full symbols with error bars) and theoretical models with (crossed symbol) and without (open symbols) inclusion of the effects of nuclear dynamics for the sidebands SB14 (black squares), SB16 (red circles) and SB18 (green diamonds) for the ionic fragments \ch{CH4+} (a,c) and \ch{CD4+} (b,d).}
  \label{Fig9}
\end{figure}

The numerical values of the complex beta parameters extracted from the experimental data (Experiment) 
are reported in Table~\ref{Tab1}. We report in the same table also those determined directly from the theoretical model including the nuclear motion (Theory (nucl. motion)) and considering fixed nuclei (Theory (fixed nuclei)). The comparison between the experimental and theoretical results is shown in the complex plane representation in Fig.~\ref{Fig9}. In general, the agreement is satisfactory. Some discrepancies are observed, in particular for the imaginary parts of the $\tilde\beta_4$ parameters in \ch{CH4+} and, to a less extent, in \ch{CD4+}.
Moreover, the values extracted from the theoretical models with and without inclusion of nuclear effects present some differences, which are more pronounced for the SB14, as also shown in Fig.~\ref{Fig3}.
The isotopic dependence of the real and imaginary parts of the asymmetry parameters derived from the simulation is very limited, while the corresponding experimental quantities are compatible within the error bars, except for the imaginary part of $\tilde\beta_4$ for the SB16. The absence of a significant isotopic effect in the photoionization phases is consistent with the results presented in refs. \cite{SCIADV-Ertel-2023,NATCOMM-Gong-2023}. 

 \begin{table}[h!]
	        \centering

         \resizebox{\textwidth}{!}{
        \caption{Comparison of the complex asymmetry parameters $\tilde{\beta}_2$ and $\tilde{\beta}_4$ for SB14-18 of \ch{CH4+} and \ch{CD4+}. The experimental parameters (first lines) were retrieved from the analytical fit of $A_{2\omega}(\theta)$ and phase $\Delta\varphi(\theta)$ according to Eqs.~(\ref{Eq_fitting1},\ref{Eq_fitting2}). 
        The theory values (second line) are at the center of the \ch{CH4+}/\ch{CD4+}-associated band (see Fig.~\ref{Fig3}), with the nuclear motion treated as described in ref.~\cite{SCIADV-Ertel-2023}. Fixed-nuclei results (third lines) are for the single geometry of the ground state neutral \ch{CH4} molecule and for the effective $I_{p}$ of 13.6~eV.}\vspace{6pt}
\label{tab:CH4+_CD4+_SB14_asymmetry_parameters}
	        \begin{tabu}{|c|c|c|c|c|}\hline
	           \textbf{$\tilde{\beta_L}$}& \textbf{SB} & \textbf{Data} & \textbf{\ch{CH4+}} & \textbf{\ch{CD4+}} \\\hline
	          \multirow{4}{*}{$\tilde{\beta}_2$} & \multirow{4}{*}{14}  & Experiment & $(2.49\pm0.21) + (0.45\pm0.14)i$ & $(2.48\pm0.19) + (0.40\pm0.11)i$\\
               &   & Theory (nucl. motion) & $ 2.47 + 0.84i$ & $ 2.47 + 0.77i$ \\
	           && Theory (fixed nuclei) & $2.63 + 0.36i$ & $2.63 + 0.36i$\\\hline
	          \multirow{4}{*}{$\tilde{\beta}_4$} & \multirow{4}{*}{14}  &  Experiment & $(0.21\pm0.44) + (0.05\pm0.08)i$ & $(0.39\pm0.35) + (0.16\pm0.06)i$\\
             &     & Theory (nucl. motion) & $ 0.31 + 0.39i$ & $0.31 + 0.37i $ \\
	           && Theory (fixed nuclei) & $0.62 + 0.24i$ & $0.62 + 0.24i$\\\hline\hline
	           %
	          \multirow{4}{*}{$\tilde{\beta}_2$} & \multirow{4}{*}{16} & Experiment & $(2.22\pm0.36) + (0.20\pm0.08)i$ & $(2.53\pm0.10) + (0.30\pm0.09)i$\\
              && Theory (nucl. motion) & $ 2.57 + 0.24i$ & $ 2.58 + 0.24i $ \\
	           && Theory (fixed nuclei) & $2.66 + 0.15i$ & $2.66 + 0.15i$\\\hline
	          \multirow{4}{*}{$\tilde{\beta}_4$} & \multirow{4}{*}{16}& Experiment & $(0.34\pm0.53) + (0.02\pm0.05)i$ & $(0.04\pm0.10) + (0.14\pm0.06)i$\\
              && Theory (nucl. motion) & $ 0.61 + 0.18i$ & $ 0.63 + 0.18i $ \\
	           && Theory (fixed nuclei) & $0.74 + 0.13i$ & $0.74 + 0.13i$ \\\hline\hline

	           %
	          \multirow{4}{*}{$\tilde{\beta}_2$} & \multirow{4}{*}{18} & Experiment & $(2.29\pm0.24) + (0.26\pm0.09)i$ & $(2.33\pm0.09) + (0.35\pm0.06)i$\\
              && Theory (nucl. motion) & $ 2.62 + 0.10i$ & $ 2.62 + 0.10i $ \\
	           && Theory (fixed nuclei) & $2.64 + 0.08i$ & $2.64 + 0.08i$\\\hline
	          \multirow{4}{*}{$\tilde{\beta}_4$} & \multirow{4}{*}{18} & Experiment & $(0.34\pm0.37) + (-0.07\pm0.05)i$ & $(0.57\pm0.10) + (-0.01\pm0.05)i$\\
              && Theory (nucl. motion) & $ 0.74 + 0.08i$ & $ 0.75 + 0.10i $ \\
	           && Theory (fixed nuclei) & $0.77 + 0.08i$ & $0.77 + 0.08i$ \\\hline
	        \end{tabu}\label{Tab1}}
	    \end{table}

\afterpage{\clearpage}
\section{Conclusions}
We have shown that the oscillating components and the photoionization phases of a RABBIT signal originating from an ensemble of randomly oriented molecules can be described by two complex asymmetry parameters $\tilde{\beta_2}$ and $\tilde{\beta_4}$, in full analogy with the atomic case. Using advanced numerical simulations, we have demonstrated that the complex asymmetry parameters extracted from the experimental and theoretical data are in satisfactory agreement. We have derived a general formula, describing the accessible magnetic and angular momentum quantum numbers for generic polarization directions of the XUV and IR pulses.

Future developments of the theory will focus on inclusion of nuclear dynamics in the electronic continuum which is presently neglected and only nuclear dynamics in the ion is included. It is expected that nuclear dynamics in the continuum becomes important in case of resonant photoionization unlike in this case where strong resonant effects have not been observed. Development of such theory and its numerical implementation are currently underway. Similarly, applications of the present methods to larger polyatomic molecules presents a further challenge for the theory. However, disentangling in the experiments the different pathways leading to the same final ionic fragments becomes even more challenging in larger systems. The implementation of coincidence spectroscopy will be crucial in obtaining information on the molecule's photoionization and subsequent photodissociation paths.

\begin{acknowledgement}

This project has received funding from the European Union's Horizon 2020 research and innovation program under the Marie Sklodowska-Curie grant agreement no.~641789 MEDEA. D.B. and E.L. acknowledge support from the Swedish Research Council (grant 2020-06384 and grant 2020-03315, respectively). D.E. and G.S. acknowledge support and funding by the Deutsche Forschungsgemeinschaft (DFG) project International Research Training Group (IRTG) CoCo 2079 and INST 39/1079. I.M. and G.S. acknowledge financial support from the BMBF project 05K19VF1, the DFG project Research Training Group DynCAM (RTG 2717), and the Georg H. Endress Foundation. G.S. acknowledges financial support from the DFG Priority Program 1840 (QUTIF) and grant 429805582 (Project SA 3470/4-1). Z.M. and J.B. acknowledge the support of the PRIMUS (20/SCI/003) project and the Czech Science Foundation (20-15548Y). Computational resources were supplied by the project “e-Infrastruktura CZ” (e-INFRA LM2018140) provided within the program Projects of Large Research, Development and Innovations Infrastructures. This work was also supported by the Ministry of Education, Youth and Sports of the Czech Republic through the e-INFRA CZ (ID:90254).

We thank C. D. Schröter, T. Pfeifer, and R. Moshammer from the Max Planck Institute for Nuclear Physics in Heidelberg for their contribution to the development of the Reaction Microscope.

\end{acknowledgement}

\appendix
\section{Appendix A: RABBIT with arbitrary photon polarizations}

The particular form of the orientationally averaged PAD for RABBIT depends on the choice of the four photon polarizations involved, see Fig.~\ref{Fig2}. The result for the general case expressed in the spherical basis for the four photon polarizations is
\begin{eqnarray}
&&I^{(p_{A1},p_{B1}),(p_{A2},p_{B2})}=\sum_{L=0}^{4}b_{L,M}Y_{L,M}^{*}(\bm{k}),\label{eq:averaged_intf_app}\\
&&b_{L,M}=(-1)^{p_{A1}+p_{A2}}\sum_{\substack{l,m,q_{1},q_{2}\\l',m',q_{1}',q_{2}'}}\mathrm{i}^{-l+l'}e^{\mathrm{i}(\sigma_{l}-\sigma_{l'})}d^{q_{1},q_{2},(A)}_{if,l,m}\left[d^{q_{1}',q_{2}',(B)}_{if,l',m'}\right]^{*}\sum_{K_{1},K_{2}}(2K_{1}+1)(2K_{2}+1)\nonumber\\
&&(-1)^{-q_{1}-q_{2}}\tj{1}{1}{K_{1}}{-q_{1}}{q_{1}'}{M_{1}}\tj{1}{1}{K_{1}}{-p_{A1}}{p_{B1}}{M_{p_1}}\tj{1}{1}{K_{2}}{-q_{2}}{q_{2}'}{M_{2}}\tj{1}{1}{K_{2}}{-p_{A2}}{p_{B2}}{M_{p_2}}\nonumber\\
&&(-1)^{m'}\sqrt{\frac{(2l+1)(2l'+1)(2L+1)}{4\pi}}\tj{l}{l'}{L}{m}{-m'}{\mu}\tj{l}{l'}{L}{0}{0}{0}\nonumber\\
    &&\tj{K_{1}}{K_{2}}{L}{M_{p_1}}{M_{p_2}}{M}\tj{K_{1}}{K_{2}}{L}{M_{1}}{M_{2}}{\mu},\label{eq:averaged_main}
\end{eqnarray}
with
\begin{eqnarray}\label{eq:M_val}
    M = p_{B1}+p_{B2}-(p_{A1}+p_{A2}).
\end{eqnarray}
In the equations above $\bm{k}$ stands for the photoelectron momentum in the LF, $\sigma_{l}$ is the Coulomb phase and $d^{q_{1},q_{2},(A/B)}_{if,l,m}$ are the molecular-frame two-photon partial-wave dipole matrix elements.
$K_1$ and $K_2$ denote the rank of the tensor operator formed from the combination of two dipole interactions (each of rank one). $L$, which determines the angular dependence, is the final rank obtained when $K_1$ and $K_2$ are coupled.
The tensor components, given by the integers $M_{1/2}$, $M_{p_1/p_2}$, and $\mu$ are fixed by the standard selection rules for 3j coefficients. The RABBIT delays are then obtained using Eq.~\eqref{eq:intf}. This is the most general form of the orientationally averaged angle-resolved RABBIT time delay. The last equation shows that terms in the angular distribution with non-zero $M$ appear as a result of an imbalance of photon polarizations between the two arms of the interferometer. The selection rule expressed in Eq.~\eqref{eq:M_val} is also an indirect selection rule on $L$ too since $L \geq |M|$. Therefore, if $M\neq 0$ then terms with $L\leq |M|$ don't contribute and we see that by controlling polarizations of the four photons it is possible to partially disentangle the role of the individual $L,M$ contributions. E.g. in the extreme case of $M=4$ only the spherical harmonic $Y_{4,4}^{*}(\bm{k})$ contributes. Furthermore, the angular distribution will contain spherical harmonics with odd $L$ values only in chiral molecules and when at least one of the photons is circularly polarized.
This can be seen~\cite{ritchiePAD1976} by swapping the signs of all magnetic quantum numbers and noticing that the overall sign of the product of the 3j coefficients will change. Therefore, the total sum is zero unless $d^{q_{1},q_{2}}_{if,l,m} \neq d^{-q_{1},-q_{2}}_{if,l,-m}$ which happens only in chiral molecules~\cite{ritchiePAD1976}.

We will now discuss the properties of the angle-resolved time delays for different choices of the photon polarizations. The complete formulas specialized to selected combinations of photon polarizations are given in Appendix B. Here we limit ourselves to qualitative aspects of the different polarization choices.

\subsubsection{Parallel linear polarizations of the XUV and IR photons}

This is the most frequently chosen configuration corresponding to $p_{i}=0$ and the interference term specialized to this case is given by Eq.~\eqref{eq:I_linear} in Appendix B. Concretely, the 3j coefficient $\tj{1}{1}{K_{i}}{-p_{i}}{p_{i}}{0} = \tj{1}{1}{K_{i}}{0}{0}{0}$ forces $K_{i}$ to be even and therefore $\tj{K_{1}}{K_{2}}{L}{0}{0}{0}$ implies that $L$ is even too (since both $K_{1}$ and $K_{2}$ are), i.e. $L=0,2,4$ and the angular distribution has the form of Eq.~\eqref{Eq1}, where only even-order $\tilde\beta$ parameters appear which is equivalent to the atomic case~\cite{PRA-Lindroth-2021} and this result does not depend on the chirality of the system.

\subsubsection{Circular polarization and chirality}

If at least one of the photons is circularly polarized then $K_{i}$ may be odd and the selection rule $K_{1}+K_{2}+L=2n$ allows $L$ to be odd too. Therefore $\tilde\beta_{1}$ and $\tilde\beta_{3}$ components may appear. Let's assume that $L$ is odd and for simplicity set $p_{A1}=p_{B1}\neq 0$ and $p_{A2}=p_{B2}=0$. This implies the following:
\begin{enumerate}
    \item The 3j coefficient $\tj{K_{1}}{K_{2}}{L}{0}{0}{0}$ implies that $K_{1}+K_{2}$ is odd.
    \item Changing $m\rightarrow -m$ and $m'\rightarrow -m'$ does not change the sign of the 3j coefficient $\tj{l}{l'}{L}{m}{-m'}{\mu}$ but requires changing the sign of $\mu$.
    \item Changing the sign of $\mu$ implies $M_{1}\rightarrow -M_{1}$ and $M_{2}\rightarrow -M_{2}$, i.e. changing signs of all $q$s.
    \item The combined sign change upon the replacements above is $(-1)^{K_{1}+K_{2}} = -1$ due to the first point and the combined sign change of the two $q$-related 3j coefficients.
\end{enumerate}
Therefore, the whole summation reduces to zero (i.e. $\beta_{1}=\beta_{3}=0$) if the two-photon partial wave dipoles are symmetric with respect to a change of all magnetic quantum numbers
\begin{eqnarray}\label{eq:nonchiral_dip}
    d_{if,l,m}^{q_{1},q_{2}} = d_{if,l,-m}^{-q_{1},-q_{2}}.
\end{eqnarray}
As mentioned above, this is precisely the case in non-chiral molecules. In chiral molecules the summation is generally non-zero since the symmetry of the dipole matrix elements is broken as in the one-photon case~\cite{ritchiePAD1976,beaulieu2018,harvey2018}. Therefore we obtain a straightforward generalization of the one-photon photoemission circular dichroism (PECD) to multi-photon above-threshold PECD. A similar case of ($n+1$) multi-photon ionization of chiral molecules has been studied before~\cite{lehmann2013}. A mixed case of circular and linear polarization in various configurations has been studied for the case of atoms recently~\cite{sorngard2020}.

\subsubsection{Polarization-skewed measurement}

In this configuration, polarizations of the XUV and NIR photons are linear but skewed at an angle $\Theta$. This setup has been realized recently~\cite{jiang2022a}. The PAD ~\cite{reid1991} and the RABBIT delays~\cite{jiang2022a} then include spherical harmonic components with $|M| \leq 2$. The exact form of the PAD is obtained as the appropriate linear combination of the general interference term according to Eq.~\eqref{eq:averaged_intf} formally including contributions of circularly polarized photons. The appearance of odd L terms in this case follows the rules for the circular polarization above.

\section{Appendix B: orientation averaging}

In this Appendix we perform an explicit derivation of the orientationally averaged PAD, Eq.~\eqref{eq:averaged_main}, for the case of two-photon interfering pathways. For the general case of $n$-photon above-threshold ionization we refer to the result presented in our earlier work~\cite{benda2021}. Throughout we employ the spherical harmonics and rotational matrices as defined in Varshalovich~\cite{varshalovich1988}.

The final continuum wavefunction in the molecular frame is given by the expression
\begin{eqnarray}\label{eq:psi_mol}
    \ket{\Psi_{\bm{k'}}^{(-)}} = \sum_{\substack{l,m}}\mathrm{i}^{l}e^{-\mathrm{i}\sigma_{l}}Y_{l,m}^{*}(\mathbf{k'})\ket{\psi_{l,m}^{(-)}}_{M},
\end{eqnarray}
where $\sigma_{l}$ is the Coulomb phase. The Wigner rotation matrix transforms angular momentum eigenstates $\Psi_{JM}$ between the laboratory and molecular frames
\begin{eqnarray}
    \Psi_{JM}(\mathbf{r'}) &=& \sum_{M'}D^{J}_{M',M}(R)\Psi_{JM'}(\mathbf{r}),\label{eq:M2L}\\
    \Psi_{JM}(\mathbf{r}) &=& \sum_{M'}[D^{-1}(R)]^{J}_{M',M}\Psi_{JM'}(\mathbf{r}_{M}) = \sum_{M'}[D^{J}_{M,M'}(R)]^{*}\Psi_{JM'}(\mathbf{r'})\nonumber\\
    &=& \sum_{M'}(-1)^{M-M'}D^{J}_{-M,-M'}(R)\Psi_{JM'}(\mathbf{r'}), \label{eq:L2M}
\end{eqnarray}
where $\mathbf{r'}$ and  $\mathbf{r}$ stand for the molecular and laboratory frame coordinates, respectively. In the following we will suppress the argument $R$ of the rotation matrices which specifies the particular rotation of the molecular frame. Photon polarization and the detector photoelectron momentum are both fixed vectors defined in the lab-frame while the photoelectron wavefunction (and the dipole matrix elements) are calculated in the molecular frame, cf. Eq.~\eqref{eq:2pdef} and Eq.~\eqref{eq:psi_mol},
\begin{eqnarray}
    M_{\bm{k'},q_{1},q_{2}}^{(2)} &=& \sum_{\substack{l,m}}\mathrm{i}^{-l}e^{\mathrm{i}\sigma_{l}}Y_{l,m}(\bm{k'})d^{q_{1},q_{2}}_{if,l,m},\label{eq:M2_mol}\\
    d^{q_{1},q_{2}}_{if,l,m} &=& \braket{\psi_{l,m}^{(-)}|\hat{D}_{q_{2}}\hat{G}_{1}^{(+)}\hat{D}_{q_{1}}|\Psi_{i}}.
\end{eqnarray}
In the equation for $M_{\bm{k'},q_{1},q_{2}}^{(2)}$ we have explicitly included both photon polarization indices $q_{1}$ and $q_{2}$. The dipole operator for lab-frame polarization $p$ is defined as
\begin{eqnarray}
    \hat{D}_{p} = \sqrt{\frac{4\pi}{3}}\sum_{i}r_{i}Y_{1,p}(\bm{r}_{i}),
\end{eqnarray}
where the sum runs over all electrons in the system. Expressing the lab-frame spherical harmonic (photon polarization) in the molecular frame using Eq.~\eqref{eq:L2M} and inserting it into Eq.~\eqref{eq:M2_mol} we obtain the dipole matrix element for the lab-frame photon polarizations $p_{1}$ and $p_{2}$
\begin{eqnarray}
       M_{\bm{k}_{fM}i,p_{1},p_{2}}^{(2)} &=& \sum_{\substack{l,m,q_{1},q_{2}}}\mathrm{i}^{-l}e^{\mathrm{i}\sigma_{l}}Y_{l,m}(\bm{k'})d^{q_{1},q_{2}}_{if,l,m}(-1)^{p_{2}-q_{2}}D^{1}_{-p_{2},-q_{2}}(-1)^{p_{1}-q_{1}}D^{1}_{-p_{1},-q_{1}}.
\end{eqnarray}
Using this result to express the RABBIT interference term we get
\begin{eqnarray}
    &&M_{\bm{k}_{fM}i,p_{A1},p_{A2}}^{(2,A)}M_{\bm{k}_{fL}i,p_{B1},p_{B2}}^{(2,B)*} = \sum_{\substack{l,m,q_{1},q_{2}\\l',m',q_{1}',q_{2}'}}\mathrm{i}^{-l+l'}e^{\mathrm{i}(\sigma_{l}-\sigma_{l'})}Y_{l,m}(\bm{k'})Y_{l',m'}^{*}(\bm{k'})\nonumber\\
    &&d^{q_{1},q_{2},(A)}_{if,l,m}\left[d^{q_{1}',q_{2}',(B)}_{if,l',m'}\right]^{*}
    (-1)^{-q_{1}-q_{1}'}(-1)^{-q_{2}-q_{2}'}D^{1}_{-p_{A2},-q_{2}}D^{1*}_{-p_{B2},-q_{2}'}D^{1}_{-p_{A1},-q_{1}}D^{1*}_{-p_{B1},-q_{1}'}\nonumber\\
    &&(-1)^{p_{A1}+p_{A2}},\label{eq:m2start}
\end{eqnarray}
where we've included the sub- and superscripts $A$ and $B$ to differentiate between the two interferometric pathways and the corresponding photon polarizations. Next we simplify the pairs of products of the rotation matrices using the identity
\begin{eqnarray}
    D^{1}_{-p,-q}[D^{1}_{-p',-q'}]^{*} = (-1)^{-p'+q'}\sum_{K_1}(2K_1+1)\tj{1}{1}{K_1}{-q}{q'}{M}\tj{1}{1}{K_1}{-p}{p'}{M_{p}}[D^{K_1}_{M_{p},M}]^{*},
\end{eqnarray}
contract the product of the two spherical harmonics and use Eq.~\eqref{eq:M2L} to transform them into the lab-frame
\begin{eqnarray}
    Y_{l,m}(\bm{k'})Y_{l',m'}^{*}(\bm{k'}) &=& (-1)^{m'}\sum_{L}\sqrt{\frac{(2l+1)(2l'+1)(2L+1)}{4\pi}}\tj{l}{l'}{L}{m}{-m'}{\mu}\tj{l}{l'}{L}{0}{0}{0}\nonumber\\
    &\times&Y_{L,\mu}^{*}(\bm{k'}),\\
    Y_{L,\mu}^{*}(\bm{k'}) &=& \sum_{M}[D^{L}_{M,\mu}(R)]^{*}Y_{L,M}^{*}(\bm{k}).
\end{eqnarray}
Inserting those expressions into Eq.~\eqref{eq:m2start} we arrive at the expression for the lab-frame interference term for a fixed orientation of the molecule given by the rotation $R$
\begin{eqnarray}
&&M_{\bm{k}_{fM}i,p_{A1},p_{A2}}^{(2,A)}M_{\bm{k}_{fL}i,p_{B1},p_{B2}}^{(2,B)*} = \sum_{\substack{l,m,q_{1},q_{2}\\l',m',q_{1}',q_{2}'}}\mathrm{i}^{-l+l'}e^{\mathrm{i}(\sigma_{l}-\sigma_{l'})}d^{q_{1},q_{2},(A)}_{if,l,m}\left[d^{q_{1}',q_{2}',(B)}_{if,l',m'}\right]^{*}\sum_{K_{1},K_{2}}(2K_{1}+1)(2K_{2}+1)\nonumber\\
&&(-1)^{-q_{1}-q_{2}}(-1)^{p_{A1}+p_{A2}}\tj{1}{1}{K_{1}}{-q_{1}}{q_{1}'}{M_{1}}\tj{1}{1}{K_{1}}{-p_{A1}}{p_{B1}}{M_{p_1}}\tj{1}{1}{K_{2}}{-q_{2}}{q_{2}'}{M_{2}}\tj{1}{1}{K_{2}}{-p_{A2}}{p_{B2}}{M_{p_2}}\nonumber\\
&&(-1)^{m'}\sum_{L}\sqrt{\frac{(2l+1)(2l'+1)(2L+1)}{4\pi}}\tj{l}{l'}{L}{m}{-m'}{\mu}\tj{l}{l'}{L}{0}{0}{0}\nonumber\\
    &&\sum_{M}[D_{M_{p_1},M_{1}}^{K_{1}}]^{*}[D_{M_{p_2},M_{2}}^{K_{2}}]^{*}[D^{L}_{M,\mu}]^{*}Y_{L,M}^{*}(\bm{k}).
\end{eqnarray}
Orientation averaging is now performed by integrating over all molecular orientations
\begin{eqnarray}
    \frac{1}{8\pi^{2}}\int\mathrm{d}R[D_{M_{p_1},M_{1}}^{K_{1}}(R)]^{*}[D_{M_{p_2},M_{2}}^{K_{2}}(R)]^{*}[D^{L}_{M,\mu}(R)]^{*} = \tj{K_{1}}{K_{2}}{L}{M_{p_1}}{M_{p_2}}{M}\tj{K_{1}}{K_{2}}{L}{M_{1}}{M_{2}}{\mu}.
\end{eqnarray}
Therefore the final orientationally averaged expression for the interference term is
\begin{eqnarray}\label{eq:intf_general}
&&I^{(p_{A1},p_{B1}),(p_{A2},p_{B2})} = (-1)^{p_{A1}+p_{A2}}\sum_{\substack{l,m,q_{1},q_{2}\\l',m',q_{1}',q_{2}'}}\mathrm{i}^{-l+l'}e^{\mathrm{i}(\sigma_{l}-\sigma_{l'})}d^{q_{1},q_{2},(A)}_{if,l,m}\left[d^{q_{1}',q_{2}',(B)}_{if,l',m'}\right]^{*}\sum_{K_{1},K_{2}}(2K_{1}+1)(2K_{2}+1)\nonumber\\
&&(-1)^{-q_{1}-q_{2}}\tj{1}{1}{K_{1}}{-q_{1}}{q_{1}'}{M_{1}}\tj{1}{1}{K_{1}}{-p_{A1}}{p_{B1}}{M_{p_1}}\tj{1}{1}{K_{2}}{-q_{2}}{q_{2}'}{M_{2}}\tj{1}{1}{K_{2}}{-p_{A2}}{p_{B2}}{M_{p_2}}\nonumber\\
&&(-1)^{m'}\sum_{L}\sqrt{\frac{(2l+1)(2l'+1)(2L+1)}{4\pi}}\tj{l}{l'}{L}{m}{-m'}{\mu}\tj{l}{l'}{L}{0}{0}{0}\nonumber\\
    &&\tj{K_{1}}{K_{2}}{L}{M_{p_1}}{M_{p_2}}{M}\tj{K_{1}}{K_{2}}{L}{M_{1}}{M_{2}}{\mu}Y_{L,M}^{*}(\bm{k}).
\end{eqnarray}
The first 3j coefficient on the last line implies
\begin{eqnarray}
 M=-M_{p_1}-M_{p_2} = p_{B1}+p_{B2}-(p_{A1}+p_{A2}).
\end{eqnarray}
In case of identical polarizations of the first and the second photon in both arms we have $p_{A1}=p_{B1}=p_{1}$ and $p_{A2}=p_{B2}=p_{2}$ giving $M=0$ and the interference term is
\begin{eqnarray}\label{eq:I_linear}
        I^{p_{1},p_{2}}&=&(-1)^{p_{1}+p_{2}}\sum_{\substack{l,m,q_{1},q_{2}\\l',m',q_{1}',q_{2}'}}\mathrm{i}^{-l+l'}e^{\mathrm{i}(\sigma_{l}-\sigma_{l'})}d^{q_{1},q_{2},(A)}_{if,l,m}\left[d^{q_{1}',q_{2}',(B)}_{if,l',m'}\right]^{*}\sum_{K_{1},K_{2}}(2K_{1}+1)(2K_{2}+1)\nonumber\\
&\times&(-1)^{-q_{1}-q_{2}}\tj{1}{1}{K_{1}}{-q_{1}}{q_{1}'}{M_{1}}\tj{1}{1}{K_{1}}{-p_{1}}{p_{1}}{0}\tj{1}{1}{K_{2}}{-q_{2}}{q_{2}'}{M_{2}}\tj{1}{1}{K_{2}}{-p_{2}}{p_{2}}{0}\nonumber\\
&\times&(-1)^{m'}\sum_{L}\sqrt{\frac{(2l+1)(2l'+1)(2L+1)}{4\pi}}\tj{l}{l'}{L}{m}{-m'}{\mu}\tj{l}{l'}{L}{0}{0}{0}\nonumber\\
    &\times&\tj{K_{1}}{K_{2}}{L}{0}{0}{0}\tj{K_{1}}{K_{2}}{L}{M_{1}}{M_{2}}{\mu}Y_{L,0}^{*}(\bm{k}),\label{eq:averaged}
\end{eqnarray}
which shows that the angle-resolved RABBIT delays contain contributions of spherical harmonics with only $M=0$ and obviously $L=0,\dots,4$. This is equivalent to the result in atoms~\cite{PRA-Lindroth-2021}. Expressing the spherical harmonic using Legendre polynomials the equation above reduces to
\begin{eqnarray}
    &&I^{p_{1},p_{2}} = \frac{1}{4\pi}\sum_{L=0}^{4}b_{L}^{p_{1},p_{2}}P_{L}(\cos\theta),\\
    &&b_{L}^{p_{1},p_{2}}=(2L+1)\sum_{\substack{l,m,q_{1},q_{2}\\l',m',q_{1}',q_{2}'}}\mathrm{i}^{-l+l'}e^{\mathrm{i}(\sigma_{l}-\sigma_{l'})}d^{q_{1},q_{2},(A)}_{if,l,m}\left[d^{q_{1}',q_{2}',(B)}_{if,l,m}\right]^{*}\sum_{K_{1},K_{2}}(2K_{1}+1)(2K_{2}+1)\label{eq:bl_general}\\
&\times&(-1)^{-q_{1}-q_{2}}(-1)^{p_{1}+p_{2}}\tj{1}{1}{K_{1}}{-q_{1}}{q_{1}'}{M_{1}}\tj{1}{1}{K_{1}}{-p_{1}}{p_{1}}{0}\tj{1}{1}{K_{2}}{-q_{2}}{q_{2}'}{M_{2}}\tj{1}{1}{K_{2}}{-p_{2}}{p_{2}}{0}\nonumber\\
&\times&(-1)^{m'}\sum_{L}\sqrt{(2l+1)(2l'+1)}\tj{l}{l'}{L}{m}{-m'}{\mu}\tj{l}{l'}{L}{0}{0}{0}\tj{K_{1}}{K_{2}}{L}{0}{0}{0}\tj{K_{1}}{K_{2}}{L}{M_{1}}{M_{2}}{\mu},\nonumber
\end{eqnarray}
where $b_{L}$ is generally a complex number since the dipole matrix elements are. This result is equivalent to Eq~(66) from our previous work~\cite{benda2021} for the case of direct multi-photon ionization and to the result derived in~\cite{demekhin2012} for the case of RABBIT with equal polarization of both photons ($p_{1}=p_{2}=p$). The complex $\tilde{\beta_{L}}$ parameters are calculated from Eq.~\eqref{eq:bl_general} as
\begin{eqnarray}\label{eq:complexbeta}
    \tilde{\beta_{L}} = \frac{b_{L}}{b_{0}}.
\end{eqnarray}

Finally, in all cases above contributions with odd $L$ appear only in chiral molecules when at least one circularly polarized photon participates~\cite{ritchiePAD1976}.

%
%
%

\bibliography{achemso-demo}

\end{document}